\documentstyle[12pt,aaspp4,epsfig]{article}

\def\bm#1{\hbox{\boldmath $#1$}}

\begin{document}

\begin{center}\uppercase{{\sc Non Equilibrium Thermodynamics and Cosmological
Pancakes Formation}}
\end{center}\vspace{0.1cm}

\author{\sc Romain Teyssier, Jean-Pierre Chi\`eze}
\affil{Centre d'Etudes de Saclay, DAPNIA, Service d'Astrophysique,
 91191, Gif sur Yvette, France.}\vspace{0.1cm}

\author{\sc Jean-Michel Alimi}
\affil{Laboratoire d'Astrophysique Extragalactique et de Cosmologie,
 CNRS URA 173, Observatoire de Paris-Meudon, 92195-Meudon, France.}
\vspace{0.1cm}

\begin{abstract}
We investigate the influence of non equilibrium thermodynamics on
cosmological structure formation. In this paper, we consider the
collapse of planar perturbations usually called ``Zel'dovich
pancakes''. We have developed for that purpose a new two fluids
(gas and dark matter) hydrodynamical code, with three different
thermodynamical species: electrons, ions and neutral particles
$T_e\ne T_i \ne T_n$). We describe in details the complex structure
of accretion shock waves. We include several relevant processes for
a low density, high temperature, collisional plasma such as
non-equilibrium chemical reactions, cooling, shock heating, thermal
energy equipartition between electrons, ions and neutral particles
and electronic conduction. We find two different regions in the
pancake structure: a thermal precursor ahead of the compression
front and an equipartition wave after the compression front where
electrons and ions temperatures differ significantly. This complex
structure may have two interesting consequences: pre-heating of
unshocked regions in the vicinity of massive X-ray clusters and
ions and electrons temperatures differences in the outer regions of
X-rays clusters.
\end{abstract}

\vspace{1mm}


\vspace{1cm}

\section{INTRODUCTION.}

The thermodynamical state of the Intergalactic Medium is of primary
importance to study the formation and the evolution of cosmic
structures. In the dense central regions of galaxy clusters,
cooling is likely to play a dominant role. In the outer regions,
the density is rather low and allows an adiabatic treatment of gas
dynamics. In the same time, non-equilibrium thermodynamics occurs
in this in this hot and diffuse plasma. Indeed, for high
temperatures ($T \simeq 10^{8}~K$) and low densities ($n_e \simeq
10^{-4}~cm^{-3}$), typical values found in the outer regions of
large X-rays clusters (Markevitch et al. 1996), the time-scale for
electrons and ions to reach thermodynamical equilibrium through
Coulomb collisions is about $t_{ei} \simeq 4 \times 10^{9}~yr$,
comparable to the Hubble time.

In order to study these low density regions where strong departure
from thermodynamical equilibrium is expected, it has been pointed
out by Kang et al. (1994) that an Eulerian code is well suited. In
this paper, we therefore present a new Eulerian code and then use
it to model non equilibrium processes during the formation of
Zel'dovich pancakes.

Pancakes appeared for a long time as fundamental tools in
Cosmology. Zel'dovich (1970) was the first to point out that
sheet-like structures could form through gravitational instability.
These pancakes was first motivated by the neutrino-dominated
scenario of structures formation, where they form naturally.
Although this scenario seems today in difficulties with
observations, walls and filaments are still observed in both
observational and theoretical studies based on other more popular
scenarios like Cold Dark Matter or Mixed Dark Matter
(\cite{CenOstriker92}, \cite{Peebles93}). Therefore, pancake
geometry is not only an idealized case for testing numerical codes,
but is also cosmologically relevant.

Hydrodynamics of pancake collapse have been studied by several
authors (\cite{Bond84}, \cite{Shapiro85}, \cite{Anninos94}) with
the baryon component, obeying the hydrodynamics equations and
coupled to collisionless dark matter particles. Different numerical
schemes for modeling the hydrodynamical equations were used. Bond
{\it et al}. (1984) and Anninos and Norman (1994) used an Eulerian
scheme, while Shapiro and Struck-Marcell (1985) used a Lagrangian
scheme. All these studies were dedicated to the calculation of the
cooled mass fraction formed via pancake collapse. On the contrary
to the previous studies, we focus in this paper on non-equilibrium
phenomena which are enhanced in the large scale, low density part
of pancakes. We introduce three temperatures to describe the
thermodynamic evolution of electrons, ions and neutral particles.
Chemical evolution of the primordial Hydrogen-Helium gas is solved
without assuming ionization-recombination equilibrium. Moreover, we
also model electronic conduction with a flux-limited diffusion
scheme.

This paper is organized as follows. In section 2 we present the
basics equations governing the system and we describe our
hydrodynamical code with validating tests. In section 3 we present
the results of simulations concerning the formation of pancakes. We
show that, in the general case, a complex structure forms, with a
thermal wave escaping out of the shocked region together with an
equipartition wave where electrons and ions temperatures can differ
significantly. We finally discuss in section 4 the r\^ole of
various parameters, such as the baryons density parameter $\Omega
_{B}$, the wavelength of the initial perturbation $L$ and the
pancake collapse epoch $a _{c}$.

\section{PHYSICS AND NUMERICAL METHODS}

\subsection{Basic equations}

The equations are written in comoving coordinates, through the
transformation $\bm r = a(t) \bm x$ where $a(t)$ is the expansion
factor. We take here $a$ as the time variable. We assume an
Einstein-de Sitter universe, with zero cosmological constant. The
velocity of Dark Matter particles and fluid elements are
respectively $\bm v = \frac{\hbox{D} \bm x} {\hbox{D}a}$ and $\bm u
= \frac{\hbox{D} \bm x} {\hbox{D}a}$. Dark matter particles
satisfy the equations of motion

\begin{equation}
\frac{\hbox{D}\bm v}{\hbox{D}a} = - \frac{2-\Omega/2}
 {a} \bm v -\frac{3\Omega}{2a^{2}} \bm \nabla _{x}\phi
\end{equation}

\noindent
$\Omega$ is the background density parameter, and $H$ the Hubble
constant (both parameters are time-dependent quantities). For
Lagrangian fluid elements, one has to add the pressure term

\begin{equation}
\frac{\hbox{D}\bm u}{\hbox{D}a} = - \frac{ 2-\Omega/2 } {a}
\bm u -\frac{3\Omega}{2a^{2}} \bm \nabla _{x}\phi
-\frac{1}{a^{4}H^{2}} \frac{ \bm \nabla _{x}P}{\rho _{B}}
\end{equation}

\noindent
and to consider the continuity equation

\begin{equation}
\frac{1}{\rho} \frac{D\rho}{Da} = -\left(\frac{3}{a}+\bm \nabla_x \cdot \bm u \right)
\end{equation}

\noindent
The total pressure $P=P_{e}+P_{i}+P_{n}$ is the sum of the
electrons, ions and neutral particles partial pressures. The
gravitational potential satisfies the Poisson equation

\begin{equation}
\bm \nabla ^2 _{x} \phi = \frac{\rho - \bar {\rho}}{\bar {\rho}}
 = \delta
\end{equation}

\noindent
where $\rho = \rho_{D}+\rho_{B}$ is the total mass density
and $\bar {\rho}$ is the mean background total mass density.
Throughout this paper, densities refer to proper physical
quantities and not to comoving quantities. We consider six chemical
species of number densities of each species $n_{e}$, $n_{HI}$,
$n_{HII}$, $n_{HeI}$, $n_{HeII}$ and $n_{HeIII}$. The partial
pressures are then related to the kinetic temperatures as $
P_{e}=n_{e}kT_{e} $, $ P_{i}=\left( n_{HII}+n_{HeII}+n_{HeIII}
\right) kT_{i} $ and $P_{n}=\left( n_{HI}+n_{HeI} \right) kT_{n} $.

We define the specific volume $V=1/\rho _{B}$, occupied by a unit
mass of baryons. The specific internal energy $E_{\alpha}$ for each
thermodynamical specy ($\alpha = e,i,n$) follows then the equation
of state for a mono-atomic gas ($\gamma = 5/3$)

\begin{equation}
E_{\alpha}=\frac{1}{\gamma - 1} P_{\alpha}V
\end{equation}

\noindent and satisfies the first law of thermodynamics

\begin{equation}
\label{energyequ}
\frac{\hbox{D}E _{\alpha}}{\hbox{D}a} =
-(\gamma-1)E_{\alpha}\left(\frac{3}{a} + \bm \nabla _{x} \cdot \bm u \right)
+\frac{D {\cal Q} _{\alpha} }{\hbox{D}a}
\end{equation}

\noindent
The first term in the right hand-side of equation (\ref{energyequ})
is the $PdV$ work due to expansion and comoving compression and the
last term is the net heat source per unit mass due to different
irreversible, non adiabatic processes.

\subsection{Thermodynamical processes}

The thermodynamical evolution of the plasma is treated in a
self-consistent way with its chemical evolution, without assuming
ionization - recombination equilibrium. The thermodynamical
processes modeled in our code are: shock heating for ions and
neutral particles, cooling and thermal conduction for electrons and
equipartition between electrons, ions and neutral particles.

We have used the collisional ionization rate and the radiative
recombination rate given by Cen (1992) which include correction
terms for very high temperatures. As we have to deal with three
different kinetic temperatures for the gas, the actual rates are
obtained by using the {\it reduced temperature} of the two
reactants (Draine 1980; Draine \& Katz 1986). In the case of
proton-electron interaction, this writes $T_{ei}=\left(\frac{
m_pT_e+ m_eT_i} {m_p+m_e}\right)$. In practice, if $T_e\simeq T_i$,
this implies a small correction of the order of $m_e/m_p$. But in
some extreme cases where $T_i \gg T_e$, this correction is not
negligible. We also considered classical cooling processes such as
ionization, recombination and line cooling, together with
bremsstrahlung and Compton cooling by the Cosmic Background
Radiation. The cooling rates are again those of Cen (1992),
modified using the reduced temperature. Note that cooling results
in an internal energy loss for the electrons only.

Massive particles (ions and neutral) share most of the entropy
deposition due to shock heating. The electrons are mainly heated by
energy exchange with the latter species. As a matter of fact, in a
perfect fluid, shocks are discontinuities in the flow, obeying
Rankine-Hugoniot relations. They imply that the post-shock
temperature for a given particle specy ``i'' is $T
\propto m_i {\cal D}^2$, where ${\cal D}$ is the up-wind fluid
velocity in the rest-frame of the shock front. Consequently, the
post-shock electron temperature is $m_{e}/m_{p}$, much lower than
the ions post-shock temperature and negligible compared to the
final equilibrium temperature. We therefore neglect electrons shock
heating. Shock heating will be treated in the code using the
artificial viscosity method (Richtmyer \& Von Neuman 1974), which
includes a linear and a quadratic viscous term. This can be written
in one dimension, {\it and in one dimension only}, as a viscous
pressure, that we add to the usual ions (resp. neutral particles)
thermal pressure.

\begin{equation}
P_{i,visc} = P_i\left( C_1 \epsilon + C_2 \epsilon^2 \right) ~~~
\hbox{where} ~~~
\epsilon = - \frac{a\Delta x}{c_{s,i}}
\left( 3 + a\bm \nabla_x \cdot \bm u \right)
\label{artvisc}
\end{equation}

\noindent
The equivalent energy source term entering equation
(\ref{energyequ}) for ions is given by

\begin{equation}
\frac{D{\cal Q}_{i,visc} }{Da}= -P_{i,visc} V
\left( \frac{3}{a}+\bm \nabla_x \cdot \bm u  \right)
\label{artviscenergy}
\end{equation}

\noindent
Similar equations apply for neutral particles. $C_1$ and $C_2$ are
two constants determined a posteriori by numerical tests, and
$\Delta x$ is the mesh spacing.

Electrons are heated by ions through Coulomb interactions, and by
neutral particles through short-range forces. The equipartition
rates are computed using the momentum transfer cross section of the
different interacting species. For example, the net kinetic energy
transfer rate per unit mass between electrons and protons is
(Spitzer 1962)

\begin{equation}
\frac{\hbox{D} {\cal Q}_{p \rightarrow e} }{\hbox{D}a}
 = -\frac{\hbox{D} {\cal Q}_{e \rightarrow p} }{\hbox{D}a}
 = - k\left( T_{i} - T_{e} \right)
 \left( \frac{ 4(2\pi)^{1/2}e^{4}m_{e}^{1/2}\ln{\Lambda _{ep}} }
 { m_{p}\left( kT_{ei} \right) ^{3/2} }
 \right) \frac{ n_{e}n_{p} }{ \rho_{B}aH }
\label{eqei}
\end{equation}

\noindent
where $T_{ei}$ and $\ln{\Lambda _{ep}}$ are respectively the
reduced temperature and the Coulomb logarithm of the two
interacting particles. The heat transfer rates between the other
chemical species can be expressed in a similar form (Draine 1980;
Chi\`eze, Pineau des For\^ets \& Flower 1998).

When electronic temperature gradients are present in the flow, a
net heat flux appears, written here in its ``classical'' form

\begin{equation}
\bm q_{cl} = - \frac{1}{a} \kappa_e \bm \nabla_x T_e
\end{equation}

\noindent
with conductivity coefficient (Spitzer 1962)

\begin{equation}
\kappa _{e} = 1.84 \times 10^{-5} ~\frac{ T_{e}^{5/2} }{ \ln{\Lambda} }
\end{equation}

\noindent
In the case of very high fluxes (very steep gradients {\it or} very
high temperatures and low densities), the flux saturates to a value
corresponding to a free transport of the electron internal energy
at a fraction of the electrons sound speed. We choose the
flux-limited diffusion scheme described in Cowie and McKee (1977).
The formulae we use here are the followings

\begin{equation}
\bm q_e =  \frac{\bm q_{cl}}{1+q_{cl}/q_{sat}}
\label{condelec}
\end{equation}

\noindent
where the maximum (saturated) value of the heat flux is given by

\begin{equation}
q_{sat} = 0.4 \left( \frac{2kT_e}{\pi m_e} \right)^{1/2} n_e k T_e
\end{equation}

\noindent
The heat source (or sink) which enter equation (\ref{energyequ}),
due to electronic conduction, is finally given by

\begin{equation}
\frac{D{\cal Q}_e}{Da} = -\frac{1}{a^2H} \bm \nabla_x \cdot \bm q_e
\end{equation}

\noindent
Ions are also able to transfer heat through ions conductivity, but
the conduction coefficient is reduced by the factor $\left(
m_{e}/m_{p} \right) ^{1/2}$, so this extra heat flux is much less
effective. Moreover the average kinetic velocity for the ion gas
after the shock front is lower than the shock velocity.
Consequently ions do not cross the shock front, except for a few
supra-thermal particles. The thermal flux due to ions is therefore
neglected. In the opposite, the electrons sound speed is $\left(
m_{p}/m_{e} \right)^{1/2}$ times larger than the shock wave
velocity (\cite{Zeldovich66}). The electron heat flux therefore
crosses easily the compression front and pre-heats efficiently the
cold gas ahead of the shock. Finally, ions will in turn be heated
through Coulomb energy exchange with electrons.

The capacity of electron to transport heat via conduction is
dramatically limited in presence of magnetic fields. Since no
evidence for strong magnetic fields has come to our knowledge, we
expect electronic conduction to occur at this rate. But during the
collapse, and more specifically during shock wave formation, a
strong enough magnetic field could be generated and reduce the
effect of conduction. Therefore, we made simulations with and
without electronic conduction.

\subsection{Numerical technics}

We present here our 1D hydrodynamical Eulerian code. The extension
of this code to a fully 3D hydrodynamical scheme is presented in a
companion paper (\cite{paper4}).

Our code is based on the operator splitting method with four
consecutive steps. The first step is called the {\bf Gravity step}.
It solves the Vlasov-Poisson equations for Dark Matter particles
and calculates the gravitational potential. The second step is
called the {\bf Lagrangian step}, and solves the adiabatic
Hydrodynamics equations in their Lagrangian form. The third step is
called {\bf Eulerian step}, it calculates the projected
hydrodynamics quantities on the fixed Eulerian grid from the
perturbed Lagrangian grid. The last step is called the {\bf
Dissipative step}; it computes all local dissipative processes,
cell by cell, using the densities resulting from the two previous
steps. Our code in its final version is of second order accuracy
both in time and in space. It allows great stability and
efficiency. We present here its general features. We then show
tests which demonstrate its ability to handle cosmological
simulations.

\subsection{General Presentation of the Code}

We consider a two-fluid system (Dark-Matter and Baryons). The
physical variables associated to Dark Matter particles (superscript
$j$) are $x^{j}$, the position of particle $j$, $v^{j}$, the
velocity of particle $j$. The discrete values of the flow on the
grid (superscript $i$) are $M^{i}$, the total baryons mass in cell
$i$, $N_{X}^{i}$, the total numbers of particle X in cell $i$,
$S_{e}^{i}$, $S_{i}^{i}$, $S_{n}^{i}$, respectively the total
entropy of electrons, ions and neutral particles in cell $i$,
$u_{x}^{i}$, the velocity in the x-direction of interface $i$ and
finally $r_{x}^{i}$, the position in the x-direction of interface
$i$.

Mass, particles numbers and entropies are zone-centered, while the
velocity is face-centered. This is the well-known staggered mesh
method (\cite{Stone92}). It allows better accuracy when computing
finite differences, and also reduces the number of interpolation to
calculate fluxes, which are defined at cell interfaces. The
variable $r_{x}^{i}$ is usual in pure Lagrangian schemes, it allows
to compute densities, while the mass remains constant. As we will
see below, it also improves the accuracy of both time integration
and flux interpolation.

The specific entropy $S_{\alpha}$ is defined for each
thermodynamical specy $\alpha$ as $S _{\alpha}=E_{\alpha} V^{\gamma
- 1}$. Using the energy equation (\ref{energyequ}), the
time-derivative of $S_{\alpha}$ reduces then to

\begin{equation}
\frac{\hbox{D}S_{\alpha}}{\hbox{D}a} = \frac{S_{\alpha}}{E_{\alpha}}
\frac{\hbox{D} {\cal Q}_{\alpha} }{\hbox{D}a}
\label{entropyequ}
\end{equation}

The thermodynamical evolution of baryons is computed using the
entropy equation (\ref{entropyequ}) rather than the energy equation
(\ref{energyequ}). This method does not introduce any numerical
spurious dissipation effects due to expansion or compression.
Consequently, an adiabatic flow ( $\hbox{D} {\cal Q}_{\alpha} /
\hbox{D}a=0$ ) is strictly adiabatic during the Lagrangian step.

\subsection{Gravity Step}

We evolve the Dark Matter particles using a Particle-Mesh algorithm
(\cite{Hockney81}) with a Predictor-Corrector scheme. This
time-integrator ensures both second-order time-accuracy and
variable time-stepping. Let us suppose that particles positions and
velocities are known at a given time $a$. We compute the predicted
positions at time $a + \Delta a$ with $(x^{j})^{(1)}= x^{j} +
\Delta a ~ v^{j}$. The superscript $(1)$ means that the quantity
is the predicted quantity evaluated at time $a + \Delta a$. The
Dark Matter density fields $\rho^{i}$ and $(\rho^{i})^{(1)}$ are
then evaluated with a CIC interpolation scheme. In order to solve
now the Poisson equation, we need to know also the predicted Baryon
density field $(\rho^{i}_B)^{(1)}$. This one is deduced from the
Baryon density and velocity fields at time $a$ by solving the
continuity equation. From the gravitational potential $\phi$ at
time $a$ and from its predicted quantity $(\phi)^{(1)}$, at time $a
+ \Delta a$, we deduce then the forces $F^{j}$ and $(F^{j})^{(1)}$
for the Dark Matter particles by inverse interpolation, at
respectively positions $x^{j}$ and positions $(x^{j})^{(1)}$.
Velocities and positions for Dark Matter particles are finally
updated according to the formula

\begin{equation}
\frac{ (v^{j})^{(2)} - v^{j} } { \Delta a } = -\frac{ 2-
 \Omega/2 }{a} \frac{ (v^{j})^{(2)} + v^{j}}{2}
-\frac{3\Omega}{2a^{2}} \frac{ F^{j} + (F^{j})^{(1)}} {2}
\end{equation}

\begin{equation}
\frac{ (x^{j})^{(2)} - x^{j} } { \Delta a } = \frac{ (v^{j})^{(2)}
+ v^{j}}{2}
\end{equation}

\noindent
where $a$ is now the {\it centered } expansion factor $a +\Delta
a/2$, $\Omega$ is the corresponding density parameter and
$(x^{j})^{(2)}$ and $(v^{j})^{(2)}$ are the updated position and
velocity for particle $j$ at time $a + \Delta a$.

\subsection{Lagrangian Step}

At this step we solve the adiabatic Hydrodynamics equations,
together with shock heating and electronic conduction. All the
quantities involved in these equations depend on spatial
derivatives of the flow. These ones are estimated with a
finite-difference scheme. Chemical, cooling and equipartition
processes which are purely local (i.e. they don't depend on spatial
derivatives of the flow) are solved in the Dissipative step.

The adiabatic Hydrodynamics equations for interface $i$ write for
an explicit scheme (first order: superscript (1)) as

\begin{equation}
\left( \frac{\Delta r_{x}}{\Delta a}^{i} \right)^{(1)}= u_{x}^{i}
\label{posequ}
\end{equation}

\begin{equation}
\left( \frac{\Delta u_{x}}{\Delta a}^{i}\right)^{(1)} =
-\frac{ 2-\Omega/2 } {a} u_{x}^{i}-\frac{3\Omega}{2a^{2}}
\frac{ \phi ^{i} - \phi ^{i-1} }{\Delta x}
-\frac{2\Delta x ^{2}}{a H^{2}}
\frac{ P ^{i} - P ^{i-1} } { M ^{i} + M ^{i-1} }
\label{vitequ}
\end{equation}

\noindent
where the pressure $P^{i}$ includes electron, ions and neutral
partial pressures, and the artificial viscous pressures for shock
heating (equation \ref{artvisc}). The entropy equations due to
electronic conduction for electrons, and to shocks heating for ions
and neutral particles for the cell $i$ are derived in a similar way
by finite differencing equations (\ref{artviscenergy}),
(\ref{condelec}) and (\ref{entropyequ}). Only the variables
($r_{x}^{i}$, $u_{x}^{i}$, $S_{e}^{i}$, $S_{i}^{i}$, $S_{n}^{i}$,
$i=1,N$) are coupled, the total mass and total numbers of particles
remains strictly constant during the Lagrangian step. In order to
integrate the previous equations, we have to compute the increments
(designed by $\Delta$) of each variables between time $a$ and time
$a+\Delta a$. Several methods are possible. The most
straightforward one is to use directly the explicit estimation we
already mentioned (first order method; see \cite{Stone92}). It
however needs a rather strong condition on the time step in order
to be stable, namely the Courant condition. The implicit method, on
the other hand, allows much larger time-steps. It is very stable,
but very CPU-time consuming, because it implies to invert a band
matrix. In this paper, we have preferred to use a second-order time
integrator scheme inspired by the implicit method. It consists to
derive second order increments by Taylor expanding the first order
increments given by equations (\ref{posequ}) and (\ref{vitequ}),
with respect to the flow variables, namely $r_x$, $u_x$, $S_e$,
$S_i$ and $S_n$. For example , the acceleration terms for interface
$i$, which depends on $u_{x}^{i-1}$, $r_{x}^{i-1}$, $S_{e}^{i-1}$,
$S_{i}^{i-1}$, $S_{n}^{i-1}$, $u_{x}^{i}$, $r_{x}^{i}$,
$S_{e}^{i}$, $S_{i}^{i}$, $S_{n}^{i}$, $u_{x}^{i+1}$, and
$r_{x}^{i+1}$ yields the following second order (superscript (2))
estimation of the velocity increment

\begin{eqnarray}
\left( \frac{\Delta u_{x}}{\Delta a}^{i}\right)^{(2)} =
\left( \frac{\Delta u_{x}}{\Delta a}^{i}\right)^{(1)}
+ \frac{\Delta a}{2}
\left[\left( \frac{\Delta r_{x}}{\Delta a}^{i-1} \right)^{(1)}
\frac{\partial}{\partial r_{x}^{i-1}}
\left( \frac{\Delta u_{x}}{\Delta a}^{i}\right)^{(1)} \right]
\nonumber
\\
\label{incvit}
~~~+ \frac{\Delta a}{2}
\left[\left( \frac{\Delta u_{x}}{\Delta a}^{i-1} \right)^{(1)}
\frac{\partial}{\partial u_{x}^{i-1}}
\left( \frac{\Delta u_{x}}{\Delta a}^{i}\right)^{(1)} \right]
\\
\nonumber
+ \cdot \cdot
\cdot~~~~~~~~~~~~~~~~~~~~~~~~~~~~~~~~~~~~~~~~~~~~
\end{eqnarray}

\noindent
where only the partial derivative with respect to the $r_{x}^{i-1}$
and $u_{x}^{i-1}$ terms has been written for sake of simplicity. It
is important to add partial derivatives of the first order
increment with respect to all variables involved in equation
(\ref{vitequ}). This method differs from the standard
``predictor-corrector'' scheme, since the second order correction
presented here is computed analytically using  partial derivatives
of the first order increments. At the end of the Lagrangian step,
we have the new entropies, the new velocities and the new interface
positions. Mass and numbers of particles have not been modified.

Finally the time-step is controlled using usual methods (see e.g.
\cite{Stone92}), with a constraint for each of the following
processes: artificial viscosity, conduction, gravity and gas
dynamics. For example, for a pure gas dynamics problem, the
time-step is controlled by the following criterion

\begin{equation}
\Delta a = C_0\min \left(
\frac{a^2H \Delta x}{\sqrt { (a^2Hu)^2+c_s^2+c_{visc}^2}}\right)
\end{equation}

\noindent
where $c_s$ is the adiabatic sound speed, and $c_{visc}$ is related
to the viscous pressure by the formula

\begin{equation}
c_{visc} = 4 \sqrt {\frac{\gamma P_{visc}}{\rho_B}}
\end{equation}

\noindent
The factor 4 in the last equation corresponds to the stability
criterion for a viscous fluid with constant viscosity coefficient
$\nu$; the equations of motion in this case are analogous to the
well documented diffusion equation, for which the stability
criterion is established as $\Delta t \le (\Delta x)^2/(4\nu)$ (see
Stone \& Norman 1992). The Courant safety coefficient $C_0$ has to
be chosen less than 0.5 in order for an explicit scheme (first
order) to be stable, and even smaller (typically 0.1) to be
accurate. On the contrary, our second order scheme remains stable
and accurate, even with the ultimately large time-stepping given by
$C_0=1$ (see the tests section).

\subsection{Eulerian Step}

We now want to re-map all variables from the disturbed Lagrangian
grid to the fixed Eulerian one. This step has to conserve mass,
numbers of particles, internal energies of electrons, ions and
neutral particles, and momentum. We defined the left-centered
momentum in cell $i$ as $P_{L}^{i}= M^{i}u_{x}^{i}$ and the
right-centered momentum in cell $i$ as $P_{R}^{i} = M^{i}
u_{x}^{i+1}$, in order to deal only with zone-centered quantities.
The projection step consists then to solve the advection equation
(written here only for the mass)

\begin{equation}
\frac{\partial}{\partial t} \int \rho dV =
- \int \rho \bm u \cdot d\bm S
\end{equation}

\noindent
for each zone which has a control volume $V^{i} = (\Delta
x)^{2}\left( r_{x}^{i+1} - r_{x}^{i} \right)$. A finite difference
approximation of this integral equation is

\begin{equation}
M _{P}^{i} - M^{i} = \Delta M_{x}^{i} - \Delta M_{x}^{i+1}
\end{equation}

\noindent
where $M_{P}^{i}$ are the projected masses on the Eulerian grid and
$\Delta M_{x}^{i}$ is the advected mass through interface $i$. This
scheme is strictly conservative for the projected variables by
construction. The main problem of the advection procedure arises
when one calculates the total mass contained in the advected
volumes. To do so, we have to calculate a realistic mass
distribution within each cell, and then integrate this distribution
in the advected volume. There are basically three tractable
methods: uniform, linear and parabolic distributions. These three
methods are known as the Donor Cell method, the Van Leer method
(Van Leer 1977) and the Piecewise Parabolic Interpolation (PPI)
method (Woodward \& Colella 1985). The first one is very simple but
quite diffusive. It is first order accurate in space. The two other
methods are much less diffusive and are respectively second and
third order accurate in space. The PPI method is the most time
consuming, while Van Leer method offers a good compromise between
accuracy and efficiency (see Stone \& Norman (1992) for a general
description).

We calculate interpolation functions for mass, momentum and total
internal energy ($E=E_{e}+E_{i}+E_{n}$) only. The numbers of
particles within the cell are distributed according to the mass
distribution, and internal energies for the 3 species are
distributed according to the total internal energy distribution.
This ensures exact mass and charge conservation within each cell,
and avoids spurious decoupling between the 3 temperatures. The
entropy are then updated from the projected internal energy, and
the velocity is updated from the left-centered projected momentum,
the right-centered projected momentum and the projected mass
according to the formula

\begin{equation}
u_{x}^{i} = \frac{ P_{R}^{i-1} + P_{L}^{i} } { M^{i-1} + M^{i} }
\end{equation}

\noindent
This equation yields an exact momentum conservation. All new
Hydrodynamics variables on the Eulerian grid are now known.

\subsection{Dissipative Step}

We solve now cell by cell, all purely local dissipative processes:
chemical reactions, thermo-chemical energy exchanges, equipartition
and cooling. All these processes are described by very stiff
equations. They imply necessarily very short time-steps, which
would slow down dramatically the whole simulation and increase the
CPU-time up to prohibiting values. Consequently to avoid this
time-step catastrophe, we solve these stiff equations by using $n$
consecutive sub-steps. At each sub-step and for each cell, we
invert a $6\times 6$ matrix for chemical reactions of the $6$
involved species and a $3\times 3$ matrix for the entropies. Our
algorithm is fully vectorized and very effective, allowing high
accuracy and stability. The number of sub-steps depends on the
physical state of the cell. A similar method was used by Anninos
and Norman (1994). Such a method is not justified as soon as the
time scale of dissipative processes is much shorter than the
dynamical time scale given by the Lagrangian step. This happens in
very dense regions where cooling can be then overestimated. We
discuss how to avoid such a mistake in \cite{paper4}. In the
Eulerian low resolution case presented here, this happens only in
one or two cells at the pancake center. However, using the
prescription presented in \cite{paper4} (which consists essentially
to turn off line cooling in the central cell only), we find that
cooling at small scales has little influence on the large scale
flow.

\subsection{Numerical Tests}

\subsubsection{Advection Test}

This test was proposed by Stone \& Norman (1992) to qualify the
advection scheme only. We consider a box of length unity filled
with a gas of homogeneous density. With a resolution of
$\hbox{N}=512$ cells and periodic boundary conditions, we model the
advection at constant velocity of a single square pulse of density
10, initially sampled by 50 cells. Our results obtained with the
three different projection schemes are very close to those obtained
by Stone \& Norman (1992) at the time when the pulse has crossed
half of the computational space. The Donor cell interpolation is
dramatically diffusive, while the two other schemes reproduce the
sharp features relatively well.

\subsubsection{Shock Tube Test}

\begin{figure}
\centerline{\hbox{\epsfig{file=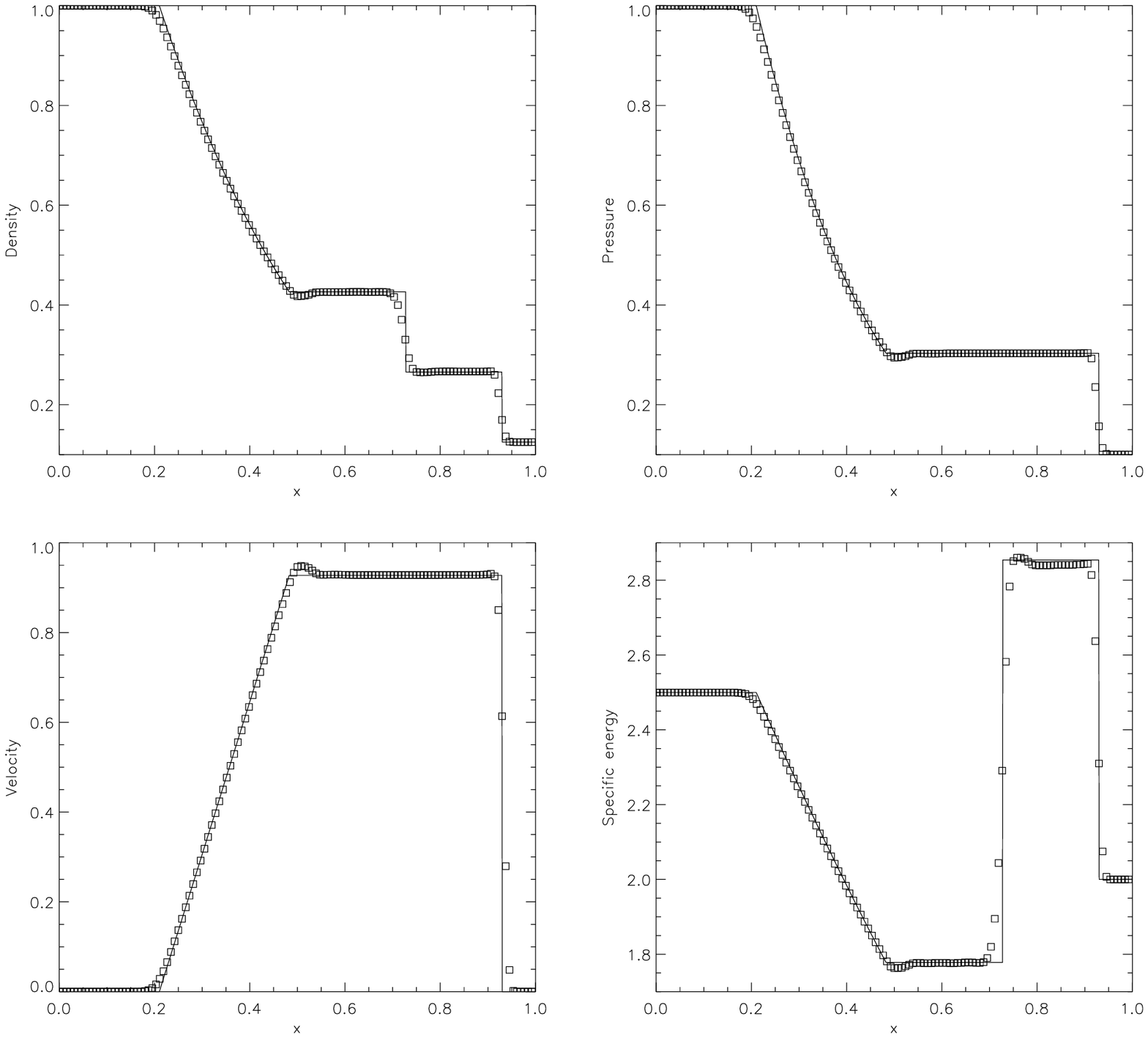,width=17cm}}}
\caption{
Density, pressure, velocity and specific energy profiles obtained
for the shock tube test at time t = 0.245. We use for that run our
standard parameters: Van Leer advection scheme, second order time
integrator, Courant safety factor $C_0=0.5$. The analytic profiles
are shown as solid lines.}
\label{tubetest}
\end{figure}

This test, also called the classical Riemann problem, is may be the
most widely used to qualify hydrodynamics codes (Sod 1978). The
initial conditions we use here are similar to those used by Stone
\& Norman (1992). We consider a box of size $\hbox{L}= 1$, $\hbox{N} =
128~\hbox{cells}$ and $\hbox{u}=0$. We separate the box in two
regions (left and right) with the following conditions $\rho_L=1$,
$p_L=1$ and $\rho_R=0.125$, $p_R=0.1$. We assume $\gamma=1.4$ and
reflective boundary conditions. We use for this test the quadratic
term of the artificial viscosity only, with $C_1=0$ and $C_2=3$. In
figure (\ref{tubetest}), we plot the different profiles obtained at
time $t=0.245$ for our standard simulation parameters, namely a
Courant safety factor $C_0=0.5$, the Van Leer advection scheme and
our second order time integrator. The results of our code is
comparable to other methods, and very close to the analytical
solution. Note however that the specific energy in the post-shock
region is slightly better reproduced here than for example in Stone
\& Norman (1992). We think that this better agreement is mainly due
to our second order time integrator.

\begin{figure}
\centerline{\hbox{\epsfig{file=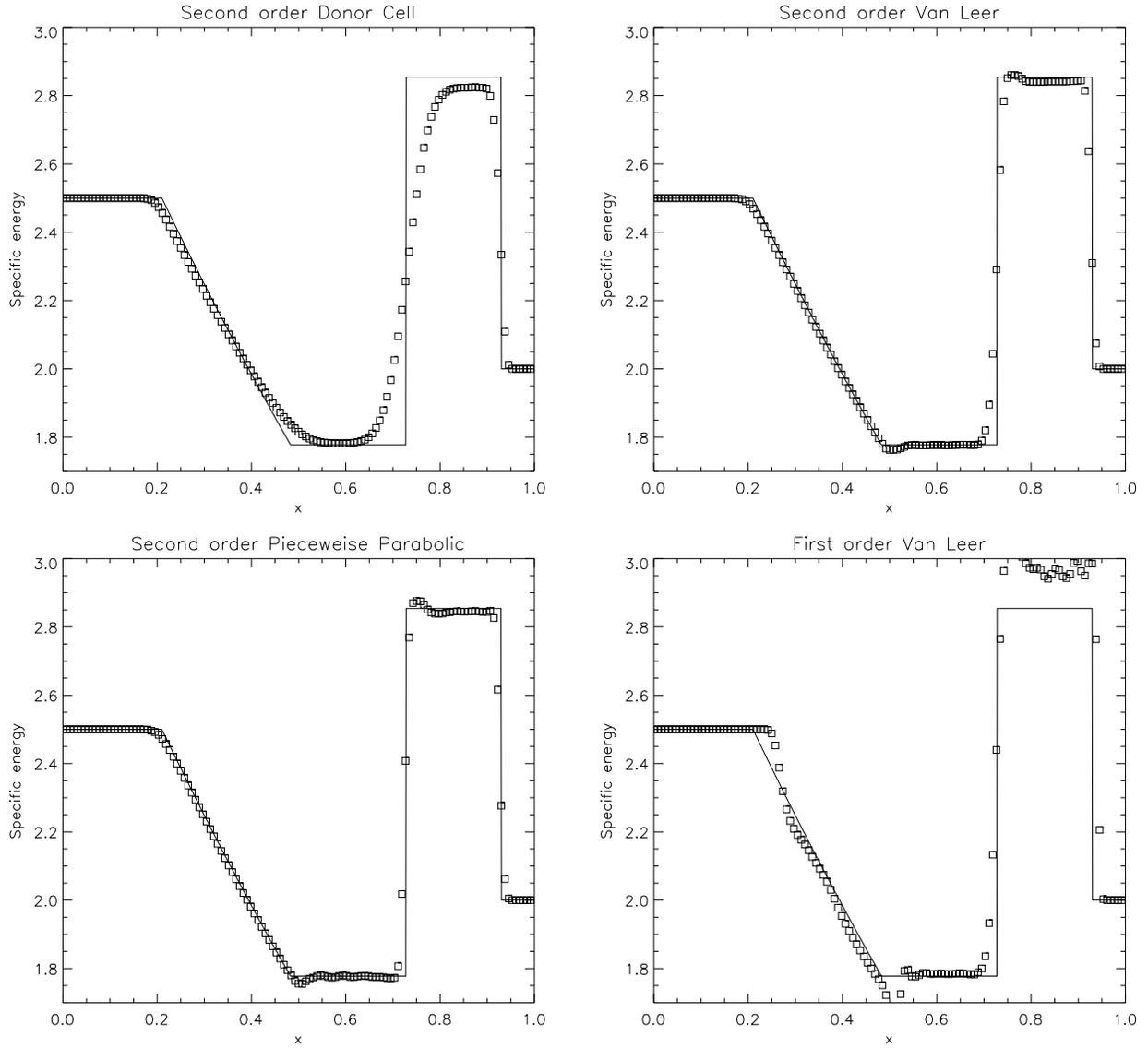,width=17cm}}}
\caption{
Specific energy profiles obtained with the shock tube test for
different advection schemes, second order time integration and the
``ultimate'' Courant safety factor $C_0=1$. We also plot the
solution obtained with the explicit method (first order) for the
same Courant safety factor. The analytic profile is shown as the
solid line in each graph.}
\label{compare}
\end{figure}

To analyze the effect of our second order time integrator, we run
the same simulation, but with a Courant safety factor $C_0=1$,
which is the ultimate possibility. We plot in figure
(\ref{compare}) the specific energy profiles obtained at time
$t=0.245$ for the three different advection schemes (Donor Cell,
Van Leer and Piecewise Parabolic) using the second order time
integrator, together with the profile obtained with the Van Leer
advection scheme, but using the first order, explicit method. Note
that in this latter case the solution is strongly unstable, while
for the three former cases, the solution is identical to the
$C_0=0.5$ case. This illustrates the interest of our second order
time integrator. We also learn from this figure that the Donor Cell
advection scheme is dramatically diffusive, and is practically of
no use. The improvement of the solution from the Van Leer to the
Piecewise Parabolic scheme is real, but not very dramatic, although
for the latter the computational cost is much higher. This justify
the use of the Van Leer method as a standard choice.

\subsubsection{Blast Waves Test}

\begin{figure}
\centerline{\hbox{\epsfig{file=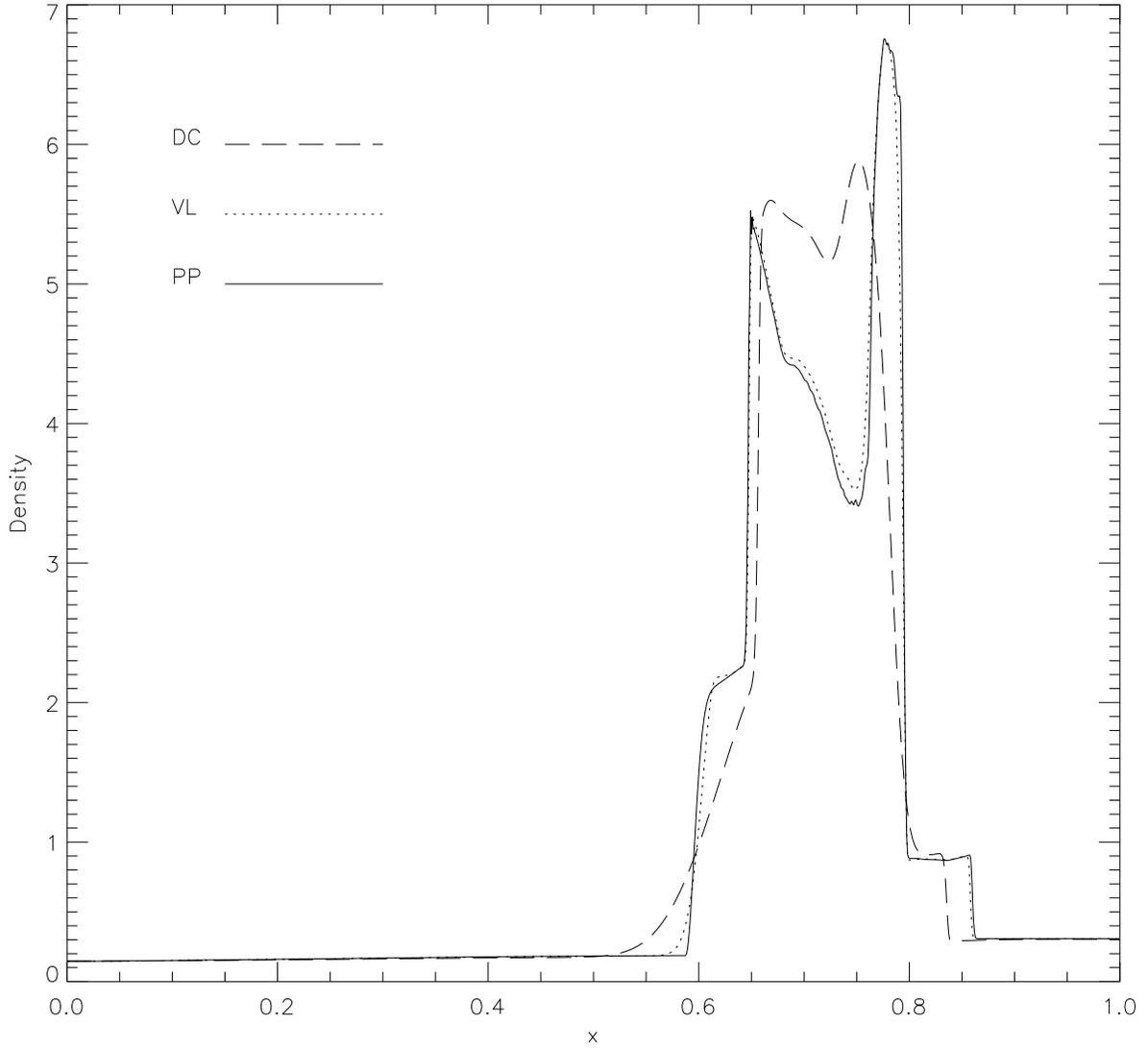,width=17cm}}}
\caption{
Density obtained with the blast waves test for different projection
schemes Donor cell (dotted line); Van Leer (dashed line); Piecewise
Parabolic (solid line). We use second order time integration and a
Courant safety factor $C_0=0.5$.}
\label{blasttest}
\end{figure}

This test was used by Woodward \& Colella (1985) to compare
different hydrodynamics codes. It is may be the most adapted test
for cosmological applications, because strong shocks are generated
with Mach numbers $\simeq 10^{5}$ and are interacting violently.
The initial conditions we use are the following (see Stone \&
Norman 1992) $\hbox{L}=1$, $\hbox{N}=1200~\hbox{cells}$, $\rho=1$
and $u=0$. In the left tenth of the box, we put $p=1000$, in the
right tenth $p=100$ and in the middle $p=0.01$. We assume
$\gamma=1.4$ and reflective boundary conditions. To have a more
precise description of the different features involved in this
simulation, see Woodward \& Colella (1985). For this test, the
linear term of the artificial viscosity is necessary to damp small
oscillations occurring otherwise after strong shock fronts (we
assume $C_1=1$ and $C_3=3$. We consider a Courant safety factor
$C_0=0.5$ and our second order time integration scheme. Had we
decided to use here the first order time integrator (explicit
method), the Courant safety factor would have been $C_0\simeq 0.1$
in order to recover similar results. We show in figure
(\ref{blasttest}) the density profiles obtained at time $t=0.038$
in order to compare with Stone \& Norman (1992) and Woodward \&
Colella (1985). At that time, the two shock waves have already
interacted at $x\simeq 0.7$ and are moving back to their original
positions. The Piecewise Parabolic scheme presents the best
agreement with Woodward \& Colella (1985) reference simulation,
although Van Leer results appears to be very similar too. As
already mentioned, Donor cell scheme is unable to handle correctly
sharp discontinuities, but more dramatic is the bad positioning of
the shock fronts. During this run, total energy conservation was
for Donor cell, Van Leer and PPI schemes respectively 7.6\%, 2.1\%
and 1.1\%.

\section{PANCAKES COLLAPSE}

\subsection{Initial Conditions}

We take for all runs the same initial conditions, beginning at
epoch $z_{i} = 200$ where matter and radiation are well decoupled.
The total density contrast has the following distribution

\begin{equation}
\delta (x) = \frac{a_{i}}{a_{c}}\cos \left(2\pi\frac{x}{L}\right)
\end{equation}

\noindent
and the baryons density distribution is taken equal to
$\rho_{B}(x)=\bar \rho_{B} \left(1+\delta (x) \right)$. Dark matter
particles are then moved according to Zel'dovich approximation,
with the displacement field corresponding to the above density
contrast. Initial velocity field for baryons is given by the linear
growing mode

\begin{equation}
u_{x} (x)= -\frac{\Omega ^{0.6}}{a} \partial _{x} \phi
\end{equation}

The temperature for ions, electrons and neutral particles are
chosen uniform and equal to $T_{e,i,n} (x) = T_{0} \left( 1+z_{i}
\right)$, where $T_{0}=2.7$ K is the cosmic radiation
background temperature today. We suppose that the helium mass
fraction is $Y=0.24$. The initial ionization fraction are taken
from Peebles (1993). In all our runs, we take $\Omega = 1$ and
$H_{0} = 50~km~s^{-1}~Mpc^{-1}$. Therefore, the single cosmological
parameter of interest here is $\Omega_{B}$. We have adopted
periodic boundary conditions to ensure total mass conservation. We
used Van Leer projection scheme since it is a good compromise
between accuracy and CPU time and a Courant safety factor
$C_0=0.5$. We consider in the following a reference pancake defined
by a comoving length $L=16~Mpc~h^{-1}$, a collapse epoch $a_c=0.2$
and a baryon density parameter $\Omega_B=0.1$.

\subsection{Adiabatic Collapse}

\begin{figure}
\centerline{\hbox{\epsfig{file=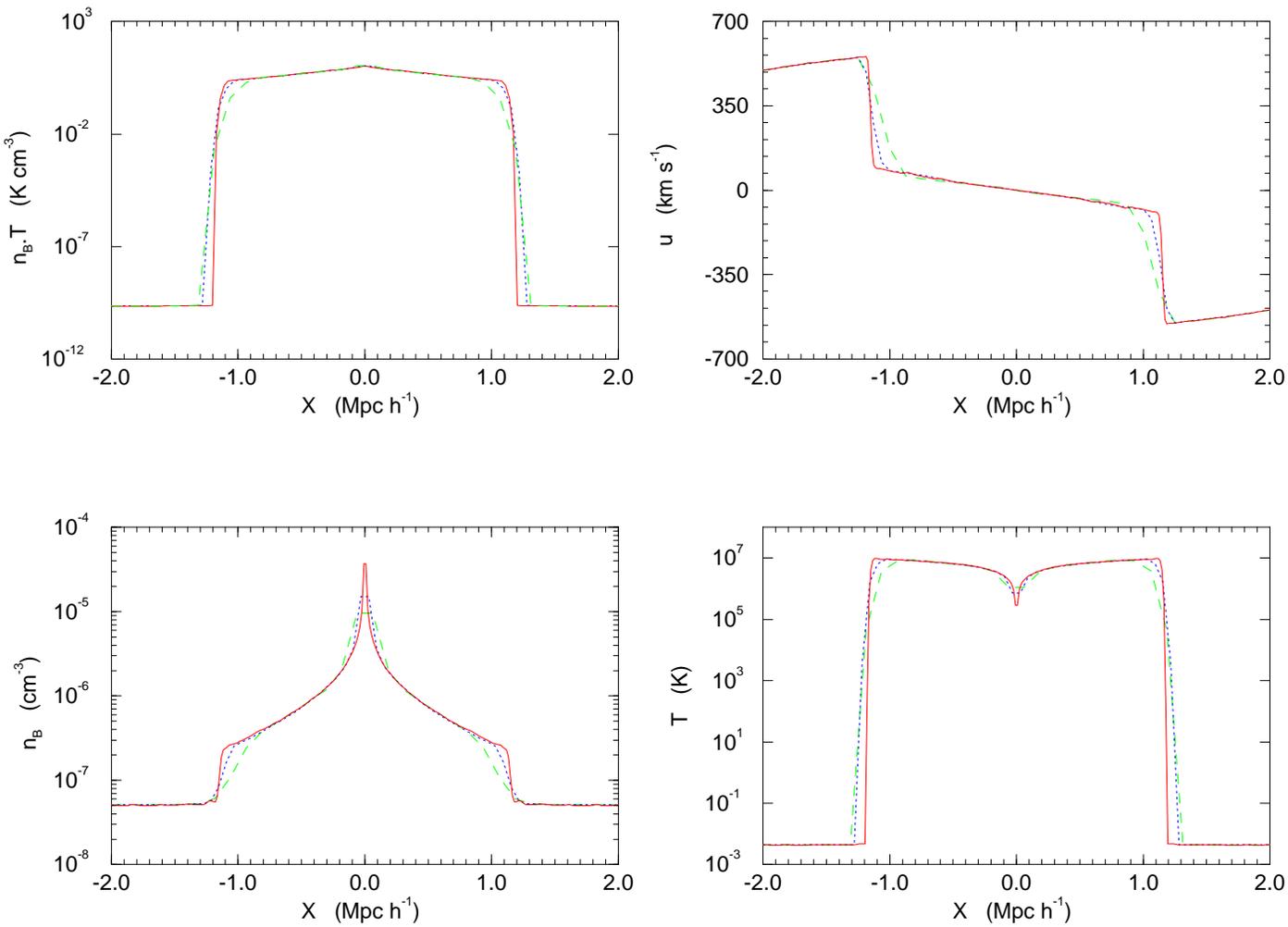,width=15cm}}}
\caption{
Density, velocity, temperature and pressure profiles at $z=0$ for
the adiabatic collapse of the reference pancake ($L=16~Mpc~h^{-1}$,
$a_c=0.2$ and $\Omega_B=0.1$. In each graph we plot three runs with
increasing resolution ($N=$128: dashed line, 256: dotted line and
1024: solid line). }
\label{adiab}
\end{figure}

In this standard case, the gas is assumed to be fully ionized, and
described by a single kinetic temperature. Sunayev \& Zel'dovich
(1972) and Bond at al. (1984) have derived analytically density and
temperature profiles in that case. They have shown that a shock
wave appears {\it off center} at a very small radius (typically a
few $10^{-6}$ Mpc) which corresponds to the sonic radius. The flow
is almost hydrostatic in this inner, unshocked region of high
density contrast. The accretion shock front propagates outward,
leaving an almost uniform pressure all over the accreted gas. Here,
contrary to Bond {\it et al.} (1984) and Shapiro \& Struck-Marcell
(1985), we don't resolve this very central region, because we are
using a regular mesh (this mesh has been chosen in such a way to
describe better the large scale, outer region of the pancake). We
are aware of the fact that the better is the resolution, the higher
is the density contrast in the central cell, until the resolution
corresponding to the sonic radius is reached. For adiabatic runs,
this has no incidence on the general aspect of the flow, since the
pressure is almost uniform, in contrast with non adiabatic runs. We
present in figure (\ref{adiab}) velocity, temperature, pressure and
baryons density profiles at $z=0$ for various mesh resolution. It
can be seen, besides an increasing sharpness of the shock front and
an increasing density contrast in the central cell, that the
resolution has little quantitative influence on the results. The
velocity field is typical of an accreting quasi-hydrostatic flow.
The usual self-similar scaling laws ($T_{ad}\propto x^{2/3}$,
$n_{ad} \propto x^{-2/3}$ and $P_{ad} \simeq const$) are well
reproduced.

\subsection{Non Adiabatic Collapse}

We turn now to the analysis of non adiabatic pancake collapse,
focusing on the temperature structure in the flow resulting from
the microscopic collisional processes among the various species. We
examine both the effects of energy exchange processes and of
electronic conduction.

\subsubsection{No Electronic Conduction}

\begin{figure}
\centerline{\hbox{\epsfig{file=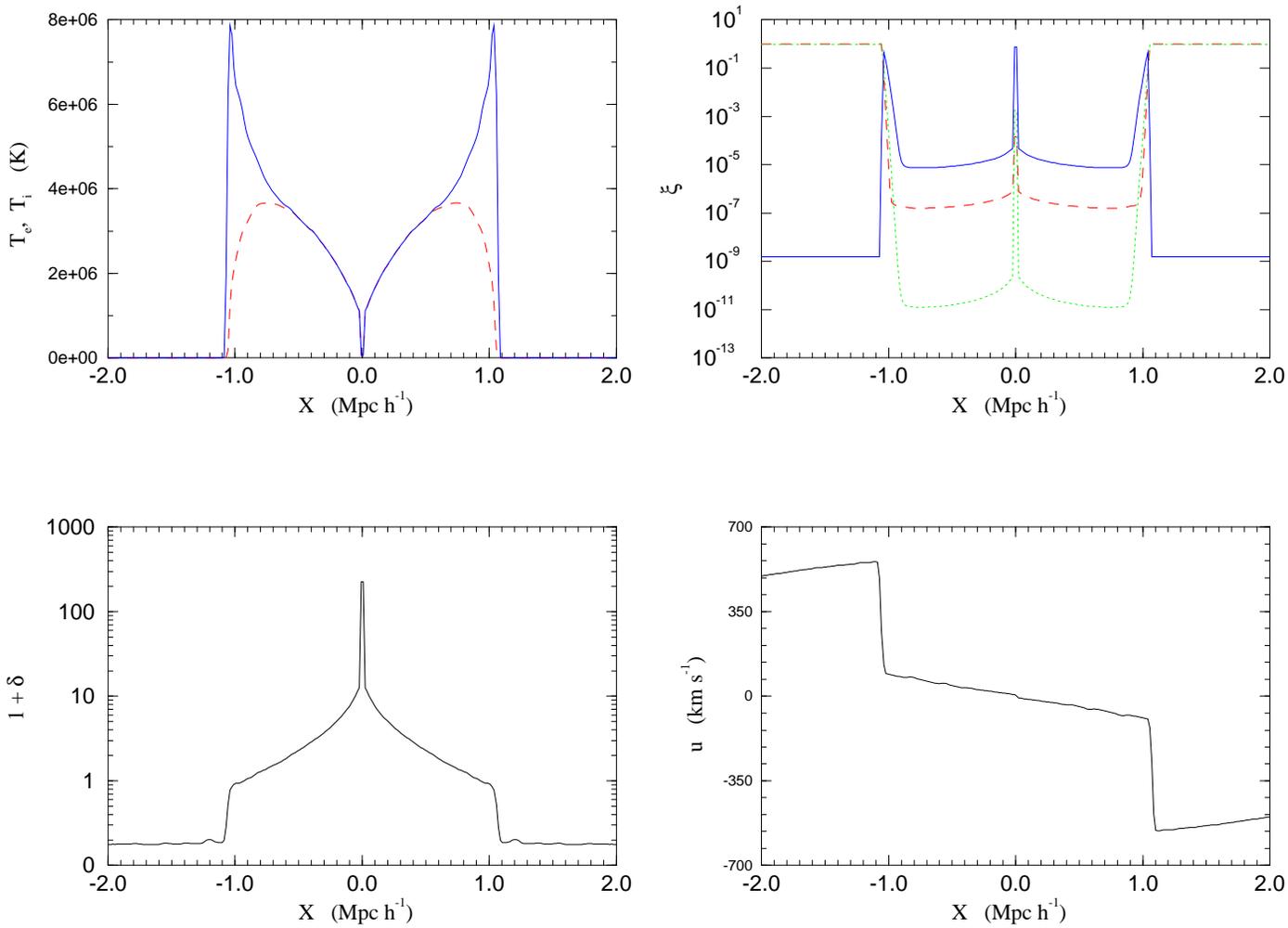,width=15cm}}}
\caption{
Gas overdensity and velocity, ions (solid line) and electrons
(dashed line) temperatures, ionization fraction for HI (dashed
line), HeI (dotted line), and HeII (solid line) at $z=0$ for the
reference run without electronic conduction. }
\label{condoff}
\end{figure}

We follow precisely chemical reactions with the corresponding
thermochemistry, collisional cooling processes and equipartition
processes, as discussed in Section 2. We first suppose that
electronic conduction is ineffective. Figure (\ref{condoff}) shows
that the electronic temperature decouples from the ion temperature
at $600~kpc~h^{-1}$ from the mid-plane. The maximum departure from
equilibrium is found near the shock front at $1.1~Mpc~h^{-1}$. In
this region, ion temperature is about $10^7$ K when electron
temperature barely reaches $10^6$ K. Furthermore, the ion and
electron temperature profiles are different, with opposite
gradients: the electron temperature steadily drops towards the
front, while the ion temperature rises.

Non equilibrium chemistry is required especially in the
post-shocked regions in which ionization gradually reaches its {\it
 near-equilibrium} value. Recall that electrons are essentially
heated behind the shock front by Coulomb collisions with hot ions.
Since equipartition processes are conservative, the total energy
density (i.e. the total pressure) is unchanged relative to the
single temperature case. By themselves, equipartition processes do
not affect the dynamics of the flow, which is only modified --
relative to the adiabatic case -- by cooling. Since the
equipartition rates per particle are proportional to density, the
temperature of ions and electrons are well coupled in the dense
parts of the pancake, but significantly decouples in the low
density outer regions. The equilibrium point (where $T_e$ and $T_i$
differ from less than 2\%) will be roughly estimated by analytical
calculations presented below.

\subsubsection{Electronic Conduction}

In that case, we suppose that electronic conduction is fully
effective, with flux-limited diffusion (Cowie \& Mc Kee 1977). We
find that conduction is saturated only in the central region, where
temperature gradients are very stiff, and in the most outer
regions, where the density is very low. Computing thermal fluxes is
a complicated task, since it depends on the ionization state of the
gas, which in turn depends on the propagation speed of the ionizing
thermal wave. We need therefore a good sampling of the ionization
front to accurately track the thermal front. This explains why we
decided to choose a linear mesh, in order to achieve a fair
sampling of temperature and abundance profiles in the thermal wave.

\begin{figure}
\centerline{\hbox{\epsfig{file=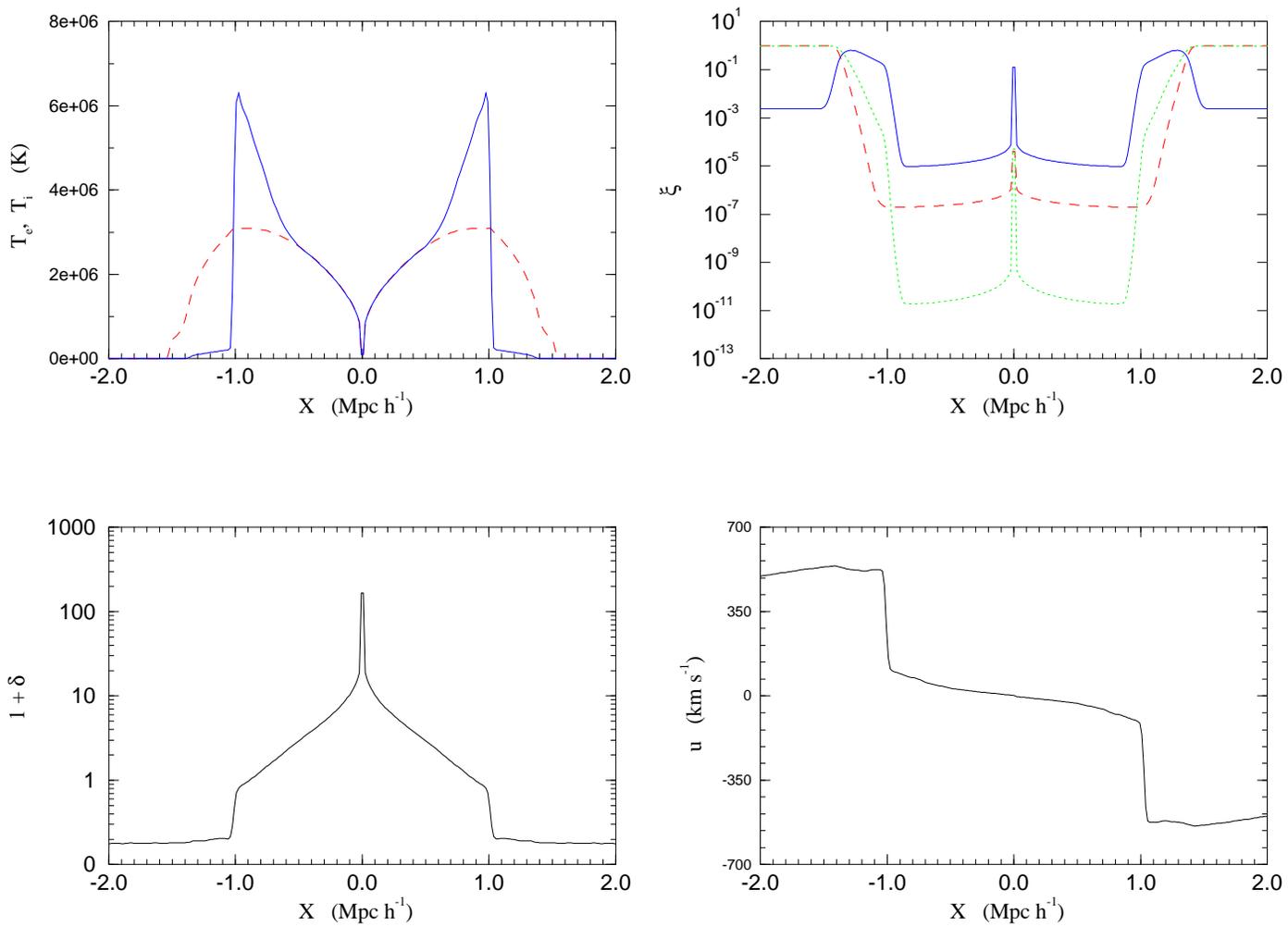,width=15cm}}}
\caption{
Same as figure (5) for the reference run with electronic
conduction. }
\label{condon}
\end{figure}

We plot in figure (\ref{condon}) the pancake state at $z=0$. Note
that the ionization state of the gas is very well sampled and that
abundances gradually evolve towards their equilibrium values. The
shock front position at $1~Mpc~h^{-1}$ is very close to its
position in the ``no conduction'' run. The main effect of
conduction is the thermal precursor, which pre-heats and
pre-ionizes the gas up to $1.5~Mpc~h^{-1}$ from the mid-plane.
Because equipartition processes are slow in this region, ion
temperature reaches only $10^5$ K, while electron temperature is
$10^6$ K. Nevertheless, this results in a slight dynamical effect
on the flow: ion pressure gradients cause a small deviation from
the pressureless velocity profile in the unshocked region, clearly
visible on this figure. Shock-heated regions occupy a total volume
of $2~Mpc~h^{-1}$, while unshocked but pre-heated ($T_i \simeq
10^{5}~K$) and pre-ionized ($T_e \simeq 10^{6}~K$) regions occupy a
total volume of $1~Mpc~h^{-1}$. As we will see in the next part,
the efficiency of electronic conduction depends strongly on the
pancake size.

\subsection{Varying the Pancake Parameters}

In this section, we study the influence of the different parameters
on the pancake structure. We develop approximate formulae which
guide us for our conclusions. We assume that, in any case, the
dynamical state of the pancake is given by the adiabatic collapse.
Following Bond {\it et al.} (1984) and Shapiro \& Struck-Marcell
(1985), it is then possible to derive interesting formulae for the
pancake evolution.

We assume first that each fluid element $q$ is shock-heated at an
epoch given by

\begin{equation}
\frac{a_s}{a_c} = \frac{\pi q}{\sin \pi q}
\label{shocktime}
\end{equation}

\noindent
This corresponds to the epoch when the corresponding collisionless
particle crosses the center of the pancake. We suppose also that
the gas is fully ionized and that the flow remains strictly
adiabatic. Before the shock front, the flow follows the
pressureless solution of pancake collapse (\cite{Zeldovich70}).
Using Rankine-Hugoniot relations and {\it assuming that the
post-shock peculiar velocity vanishes}, we get the post-shock
temperature and the post-shock overdensity

\begin{equation}
kT = \frac{1}{12}\mu m_p a^{-1} (H_0L)^{2}q^2
\label{shockT}
\end{equation}
\begin{equation}
1+\delta = 4/\left( 1-\pi q\frac{\cos \pi q}{\sin \pi q} \right)
\end{equation}

\noindent
As we assumed that the peculiar velocity is zero in the
post-shock region, the temperature evolves afterwards as $a^{-2}$
and the density as $a^{-3}$. Mass conservation implies also that
the Eulerian position of a given fluid element in the shocked
region is given by

\begin{equation}
x = \int_{0}^{q}\left( 1 - \pi q \frac{\cos \pi q}{\sin \pi q} \right)
\frac{dq}{4}
\end{equation}

\noindent
This allows us to describe the dynamical evolution (single
temperature case) of the pancake. Let us now estimate the
thermodynamical evolution using our three temperatures formalism.
The equipartition time-scale for electron-ion energy exchange is
given by

\begin{equation}
t_{ei}=503 \frac{T_{ei}^{3/2}}{n_e\ln \Lambda} \quad \hbox{sec.}
\end{equation}

Assuming that the effective temperature $T_{ei}$ is equal to the
post-shock adiabatic temperature, the equipartition time-scale
remains then constant during the post-shock evolution (we neglect
the slow variation of the Coulomb logarithm, taking
$\ln\Lambda\simeq 40$). It is now possible to solve analytically
the equipartition equation

\begin{equation}
\frac{d}{dt}\left(T_e - T_i\right) + 2\frac{\dot a}{a}\left(T_e - T_i\right)
 = -\frac{4}{t_{ei}}\left(T_e - T_i\right)
\end{equation}

The single temperature, given by equation (\ref{shockT}), is
related to $T_e$ and $T_i$ by

\begin{equation}
T = \frac{n_eT_e + n_iT_i}{n_e + n_i}
\end{equation}

\begin{figure}
\centerline{\hbox{\epsfig{file=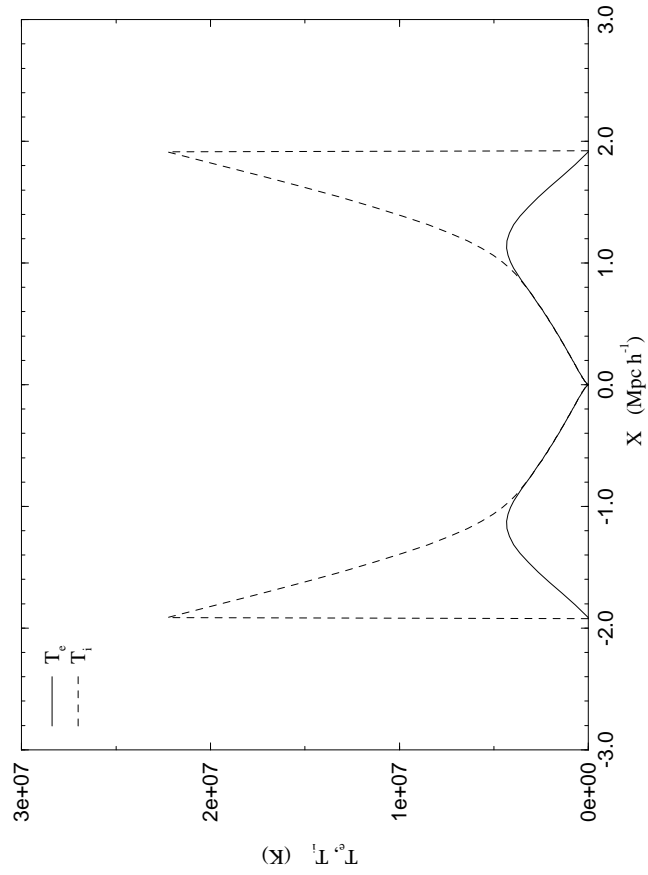,width=10cm}}}
\caption{
Analytical ions (dashed line) and electrons (solid line)
temperatures profiles for the reference pancake at $z=0$. Compare
this figure to the numerical results obtained in figure (5). }
\label{Tprofilana}
\end{figure}

We finally obtain semi-analytical temperatures profiles, plotted in
figure (\ref{Tprofilana}), for the reference pancake at $z=0$. Note
that numerical results agree qualitatively well with our analytical
theory. Because we assumed that the post-shock velocity vanishes,
we overestimate the shock front position and the post-shock
temperature. In the numerical calculation, Lagrangian fluid
elements pile up deeper than we assume in our analytical
calculation. This explain the visible differences between numerical
and analytical results.

The pancake structure is fully described by three characteristic
points, as we already presented in the upper sections: the {\it
 compression point} where ions and neutral particles are shocked, the {\it
 thermal point} which marks the end of the thermal wave and the {\it
 equilibrium point} where equilibrium between electrons and ions is
recovered. In the small $q$ limit, Bond et al. (1984) derived the
Lagrangian coordinate of the compression point

\begin{equation}
q_s \propto \left( 1 - \frac{a_c}{a}\right) ^{1/2}
\end{equation}

\noindent
We can write, assuming that $a-a_c \ll a_c$

\begin{equation}
q_s \propto \left( \frac{a-a_c}{a_c} \right) ^{1/2}
\label{shockana}
\end{equation}

\noindent
We derive here (for $q \ll 1$ and $a-a_c \ll a_c$) the
Lagrangian coordinate of the equilibrium point (where electrons and
ions temperatures differ from only 2\%)

\begin{equation}
q_{eq} \propto \left(\frac{a-a_c}{a_c}\right)^{1/5}
L^{-3/5}\Omega _{B}^{1/5}
\label{equiana}
\end{equation}

\noindent
It appears that this coordinate depends strongly on the
pancake size. This is due to the strong dependence on temperature
($T^{3/2}$) of the equipartition time-scale. When the baryon
density parameter is decreased from 0.1 to 0.01, the equilibrium
point reaches much deeper regions. The final mass at
thermodynamical equilibrium is about 85\% of the total shock-heated
mass for the $\Omega_B=0.1$ case, and decreases to about 50\% for
the $\Omega_B=0.01$ case. The time-dependence of the equilibrium
point is much slower than for the shock front. This means that
non-equilibrium features appear mainly at late epoch and in the
most outer regions where density has decreased (and temperature has
increased) sufficiently.

\begin{figure}
\centerline{\hbox{\epsfig{file=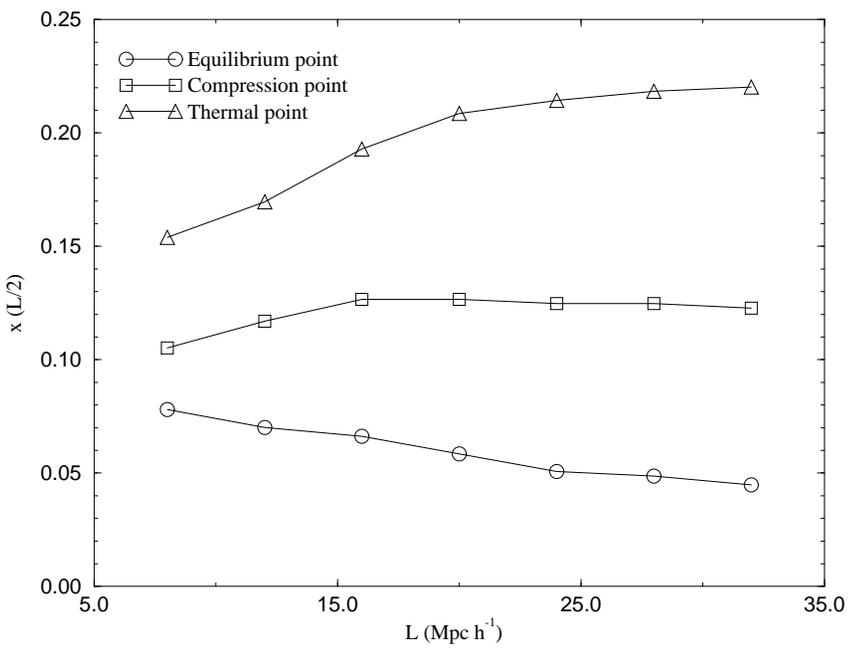,width=10cm}}}
\caption{
Eulerian comoving coordinates of the three characteristic points
(see text) which describe non-equilibrium features in the flow for
various pancake sizes. }
\label{distances}
\end{figure}

Formula (\ref{equiana}) is valid only for the small $q$ regime. To
test the validity of our description at later epoch (the large $q$
regime), we run numerical calculations with different pancake
sizes. All these runs were computed assuming that electronic
conduction is effective, and that $\Omega_B = 0.1$, $a_c=0.2$. We
plot in figure (\ref{distances}) the shock front Eulerian position
$x_s$, the equilibrium point position $x_{eq}$ and the thermal
front position $x_{th}$ obtained in these simulations. The shock
front position is almost independent of the pancake size, as
expected by formula (\ref{shockana}). The equilibrium point is
found in deeper regions as $L$ increases, as expected by formula
(\ref{equiana}). The thermal wave has approximately the symmetrical
behavior than the equipartition wave.

For $L\simeq10~Mpc~h^{-1}$, cooling starts to be important and the
flow does not remain adiabatic. The main consequence is that the
compression point propagates outwardly at a lower velocity than for
the adiabatic case. This characteristic pancake length corresponds
roughly to pancakes for which the {\it average} cooling time is
equal to the Hubble time. Indeed, assuming that the mean
temperature is given by equation (\ref{shockT}) with $q=1/4$,
$\bar\delta \simeq 10$ and using the {\it equilibrium} cooling
curve to calculate the cooling rate, we find $L_{cool}\simeq
8~Mpc~h^{-1}$.

\section{CONCLUSIONS}

In this paper, we use a three temperature formalism to describe the
thermodynamical evolution of pancakes. We show that the assumption
of thermodynamical equilibrium between ions and electrons is not
valid in the outer regions of pancakes. The corresponding
temperature profiles show differences up to one order of magnitude
near the shock front and reach thermodynamical equilibrium near the
centre, where the overdensity is $\delta \simeq 10$. Electrons and
ions decoupling is stronger for large pancake sizes or for low
$\Omega_B$. This could have observational consequences for X-ray
temperature profiles. The dynamical (total) pressure in the outer
regions of clusters can differ from the observed electrons pressure
up to one order of magnitude. This effect can be even stronger for
non-relaxed clusters, where recent mergers strongly decoupled
electrons and ions through shocks. This could lead to an
underestimate of the cluster mass in this regions. In the central
hydrostatic part of clusters, where thermodynamical equilibrium is
efficiently recovered, such effect are not likely to appear. We
address this question using a fully 3D X-ray clusters modeling in a
companion paper (Chi\`eze, Alimi \& Teyssier 1998)

In the outermost regions, we show that electronic conduction, if
effective, leads to a thermal wave escaping the shock-heated
regions of the pancake. This thermal precursor could have
interesting cosmological consequences, such as late reionization
and heating of non collapsed regions. In our case, the precursor is
strongly confined by the very high infall velocity towards the
pancake centre. In a more complex geometry, one can imagine that
the size of the precursor could be much more extended, leading to
efficient heating of the intergalactic medium. This effect might be
detected in the vicinity of very large X-rays clusters.

Non-equilibrium regions (from the thermal precursor to the
equilibrium point) are very extended and dominate {\it in volume}
the structure of pancakes (this paper) and X-ray clusters
(Chi\`eze, Alimi \& Teyssier 1998). Using high sensitivity
experiments, like the X-ray satellite XMM (Fabian 1987), we will be
able in the future to observe the low density part of clusters,
where all these processes are likely to be very important.

The authors would like to thank an anonymous referee for the
constructive remarks that allow us to increase the quality of our
work.

\pagebreak

\end{document}